\documentclass[a4paper,fleqn,usenatbib,useAMS]{mnras}
\usepackage{newtxtext,newtxmath}
\usepackage[T1]{fontenc}
\usepackage{ae,aecompl}
\usepackage{graphicx}    
\usepackage{amsmath}     
\usepackage{amssymb}     
\usepackage{multicol}    
\usepackage{bm}          
\usepackage{pdflscape}   
\newcommand{\kms}{\,km\,s$^{-1}$}
\title[Luminous type IIP SN~2013ej] 
   {Luminous type IIP SN~2013ej with high-velocity $^{56}$Ni ejecta}
\author[V. Utrobin \& N. Chugai]{
V. P. Utrobin$^{1,2}$\thanks{E-mail: utrobin@itep.ru}
and
N. N. Chugai$^{3,4}$
\\
$^{1}$Institute of Theoretical and Experimental Physics,
      B.~Cheremushkinskaya St. 25, 117218 Moscow, Russia \\
$^{2}$Max-Planck-Institut f\"ur Astrophysik, Karl-Schwarzschild-Str. 1,
      85748 Garching, Germany \\
$^{3}$Institute of Astronomy, Russian Academy of Sciences, Pyatnitskaya
      St. 48, 119017 Moscow, Russia\\
$^{4}$Sternberg Astronomical Institute, Moscow M.V. Lomonosov State University,
      Universitetskij pr., 13, 119234 Moscow, Russia
}

\date{Accepted XXX. Received YYY; in original form ZZZ}

\pubyear{2017}

\begin{document}
\label{firstpage}
\pagerange{\pageref{firstpage}--\pageref{lastpage}}
\maketitle
%
\begin{abstract}
We explore the well-observed type IIP SN 2013ej with peculiar luminosity 
   evolution.
It is found that the hydrodynamic model cannot reproduce in detail the
   bolometric luminosity at both the plateau and the radioactive tail. 
Yet the ejecta mass of $23-26\,M_{\sun}$ and the kinetic energy of
   $(1.2-1.4)\times10^{51}$\,erg are determined rather confidently.  
We suggest that the controversy revealed in hydrodynamic simulations stems
   from the strong asphericity of the $^{56}$Ni ejecta.
An analysis of the asymmetric nebular H$\alpha$ line and of the peculiar
   radioactive tail made it possible to recover parameters of the asymmetric
   bipolar $^{56}$Ni ejecta with the heavier jet residing in the rear
   hemisphere.
The inferred $^{56}$Ni mass is $0.039\,M_{\sun}$, twice as large compared to
   a straightforward estimate from the bolometric luminosity at the early
   radioactive tail. 
The bulk of ejected $^{56}$Ni has velocities in the range of $4000-6500$\kms. 
The linear polarization predicted by the model with the asymmetric ionization 
   produced by bipolar $^{56}$Ni ejecta is consistent with the observational
   value.
\end{abstract}

\section{Introduction} 
\label{sec:intro}
The luminous type IIP supernova (SN~IIP) 2013ej in M74 \citep{KZL_13} gets into
   the limelight because of its early discovery, close location
   ($D\approx9$\,Mpc), and the unusual behavior.
Specifically, at the plateau phase the flux declines more rapidly than normally
   in SNe~IIP \citep{HWZ_15}, which brought about a confusion with the
   classification of SN~2013ej, so one can meet type IIP \citep{HWZ_15},
   type IIP/IIL \citep{MDJ_17}, and type IIL \citep{BSK_15} (below we prefer
   to use type IIP/L). 
Even more amazing is the luminosity evolution at the radioactive tail:
   decline at the early radioactive tail is more rapid than the $^{56}$Co
   decay rate, which is a typical feature of SNe~IIP \citep{DKV_16}.
Moreover, the decline rate shows a slowdown after about 200\,days instead of
   a decline acceleration.   
\citet{YJV_16} attribute the rapid decline at the radioactive tail to either
   a low ejecta mass, a high explosion energy, an extreme outward mixing of
   $^{56}$Ni, or a combination of these factors; the decline slowdown is not
   discussed.
An interesting feature is the width and pronounced asymmetry of the nebular
   H$\alpha$ emission; the latter could indicate the bipolar $^{56}$Ni ejecta
   \citep{BSK_15} by analogy with SN~2004dj \citep{CFS_05}.
In this regard the detected linear polarization at the level of $\sim$1$\%$
   \citep{MDJ_17,KPEK_16} is also reminiscent of SN~2004dj.
The X-ray emission detected by {\em Chandra} and {\em Swift} indicates the
   mediocre wind density --- characteristic of a red supergiant (RSG)
   \citep{CRS_16}.

A good observational coverage of the light curve starting from the shock
   breakout and an extended set of spectra make SN~2013ej highly valuable
   object for the hydrodynamic modelling and the study of asymmetry effects
   in SNe~IIP. 
Two hydrodynamic models for SN~2013ej have been reported \citep{HWZ_15,MPV_17}. 
However, both do not fit expansion velocities at the photosphere, which poses
   a question, whether supernova parameters are reliably inferred.
The issue of adequate hydrodynamic parameters (ejecta mass, explosion energy,
   and pre-SN radius) therefore remains on the agenda.
Note that both mentioned hydrodynamic models also do not touch the issue of
   the radioactive tail.
The luminosity evolution at the radioactive tail, however, is closely related
   to the $^{56}$Ni mass and mixing and therefore should be considered as
   a crucial observational constraint for the SN~2013ej model.

Here we address the issue of hydrodynamic parameters of SN~2013ej upon the
   basis of a standard one-dimensional (1D) hydrodynamic modelling of the
   bolometric light curve and expansion velocities. 
We also consider the problem of the $^{56}$Ni distribution imprinted in the 
   nebular H$\alpha$ profile and in the luminosity decline rate at the 
   radioactive tail. 
We start with the hydrodynamic modelling assuming different possibilities 
   for the $^{56}$Ni mass and mixing (Section~\ref{sec:hydro}). 
We then consider an effect of the bipolar $^{56}$Ni jets on the nebular
   H$\alpha$ profile and the radioactive tail (Section~\ref{sec:hajets}). 
The ejecta model and the recovered $^{56}$Ni distribution are used
   for the calculation of the polarization which in turn is compared to 
   the observed polarization (Section~\ref{sec:polar}).
Results are summarized and discussed in Section~\ref{sec:disc}.

Below we adopt, following \citet{DKV_16}, the distance to SN~2013ej
   $D=9.0$\,Mpc and the reddening $E(B-V)=0.09$\,mag.

\section{Hydrodynamic modelling}
\label{sec:hydro}
%
\begin{table}
\centering
\caption{Parameters of hydrodynamic models.}
\label{tab:hydmod}
\begin{tabular}{lccccc}
\hline
\noalign{\smallskip}
Model  & $R_0$ & $M_{ej}$ & $E$ & $M_{\mathrm{Ni}}$ & $v_{\mathrm{Ni}}^{max}$ \\
       & $(R_{\sun})$ & $(M_{\sun})$ & ($10^{51}$ erg) & $(M_{\sun})$ &  km\,s$^{-1}$ \\
\noalign{\smallskip}
\hline 
\noalign{\smallskip}
m23e1.2 & 1500 & 23.1 & 1.18 & 0.020 & 5000 \\
m26e1.4 & 1500 & 26.1 & 1.40 & 0.039 & 6500 \\
\noalign{\smallskip}
\hline
\end{tabular}
\end{table}
The light curve of SN~2013ej with a luminous plateau indicates that the pre-SN
   star was a red supergiant (RSG).
Although at first glance it is advisable to explode a RSG model prepared by
   the evolutionary computations, the previous hydrodynamic simulations have
   demonstrated that the explosion of the evolutionary RSG is not able to
   describe all the essential features of the light curve
   \citep{UC_08}.
This problem in fact was first revealed for the peculiar type IIP SN~1987A. 
To produce sensible fit of the light curve, mixing at the He/H composition
   interface should be manually adjusted \citep{Woo_88}.

We therefore, as usually, choose a nonevolutionary hydrostatic RSG pre-SN
   model as the initial data for hydrodynamic simulation.
A density profile and macroscopic mixing between metal, helium, and hydrogen
   components in such a model are adjusted to fit the photometric and
   spectroscopic data.
Strictly speaking, the nonevolutionary model thus prepared is not a pre-SN
   in proper sense of this term.
This model should be thought rather as an outcome of the Rayleigh-Taylor (RT)
   mixing caused by the shock propagation following the {\em explosion}.
Yet for convenience we call this nonevolutionary model the ``pre-SN''.    
Remarkably, recent three-dimensional (3D) simulations of the SN explosion of
   a RSG star show that the RT mixing modifies both the density and composition
   gradients at the composition interfaces in accord with the nonevolutionary
   pre-SN model used formerly in 1D simulations of the normal type IIP
   SN~1999em \citep{UWJM_17}.

\begin{figure}
   \includegraphics[width=\columnwidth, clip, trim=0 237 41 138]{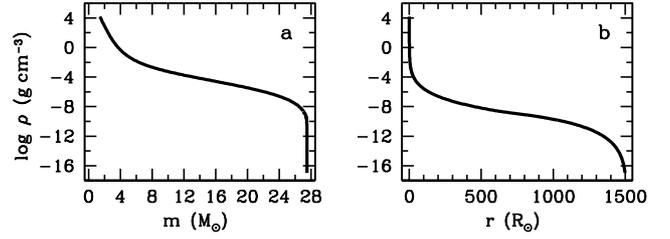}
   \caption{%
   Density distribution as a function of interior mass (Panel \textbf{a}) and
      radius (Panel \textbf{b}) for the pre-SN model m26e1.4.
   The central core of 1.4\,$M_{\sun}$ is omitted.
   }
   \label{fig:denmr}
\end{figure}

\begin{figure}
   \includegraphics[width=\columnwidth, clip, trim=7 17 44 249]{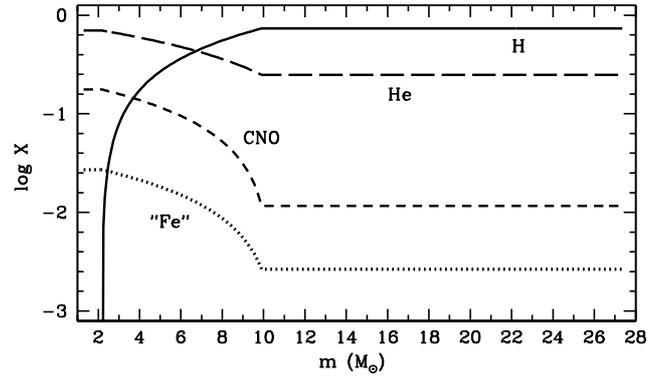}
   \caption{%
   Mass fraction of hydrogen (\emph{solid line\/}), helium
      (\emph{long dashed line\/}), CNO elements (\emph{short dashed line\/}),
      and Fe-peak elements excluding radioactive $^{56}$Ni
      (\emph{dotted line\/}) in the ejecta of the pre-SN model m26e1.4.
   }
   \label{fig:chcom}
\end{figure}

The explosion simulation of SN~2013ej utilizes the 1D hydrodynamic code with
   the radiation transfer \citep{Utr_04}. 
The explosion is initiated by a supersonic piston applied to the bottom of
   the stellar envelope at the boundary with the collapsing 1.4\,$M_{\sun}$
   core.
As a result of an extended simulations we come to alternative models
   m23e1.2 and m26e1.4 (Table~\ref{tab:hydmod}) with two options of
   the $^{56}$Ni mass and mixing extent. 
The density and composition distributions in the pre-SN model m26e1.4
   (called fiducial) are shown in Figs.~\ref{fig:denmr} and \ref{fig:chcom},
   respectively.
The pre-SN model m23e1.2 has similar density and composition distributions
   except for $^{56}$Ni: here, unlike model m26e1.4, the $^{56}$Ni abundance
   is assumed constant by mass, which means that the $^{56}$Ni is center
   concentrated.
In both models the pre-SN radius is 1500\,$R_{\sun}$, while the ejecta mass
   and the explosion energy are different. 

\begin{figure}
   \includegraphics[width=\columnwidth, clip, trim=0 18 52 249]{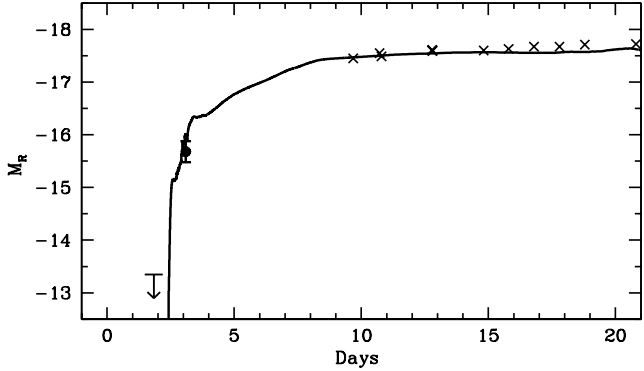}
   \caption[]{%
   $R$-band light curve (\emph{solid line\/}) during the first 20 days for
      the fiducial model m26e1.4.
   Arrow marks the upper limit $V>16.7$ for a non-detection on July 23.54
      \citep{SKS_13}, shown here for reference.
   Filled circle is the first observation of SN~2013ej in $R$-band with
      the uncertainty of about 0.2\,mag announced by \citet{LLW_13}.
   Crosses are the observational points of \citet{DKV_16}.
   The time of the shock breakout of model m26e1.4 is identified with
      the epoch of the SN detection on July 24.125 reported by C. Feliciano.
   }
   \label{fig:onset}
\end{figure}

The early detection of SN~2013ej provides an opportunity to verify the fiducial
   model m26e1.4 by comparing the calculated $R$-band light curve with
   the observations at the shock breakout stage (Fig.~\ref{fig:onset}).
The model $R$-band light curve fits the data fairly well including 
   the first $R$-band observation of SN~2013ej \citep{LLW_13} and 
   the $R$-band photometry of \citet{DKV_16}.
Identifying the initial steep jump of the $R$-band light curve at the
   shock breakout with a non-photometric detection of SN~2013ej on July 24.125,
   which was reported by C. Feliciano on the Bright Supernovae website%
\footnote{\url{http://www.rochesterastronomy.org/supernova.html}},
   fixes the explosion at MJD=56494.705 (July 21.705), i.e.,
   2.195\,days earlier compared to the estimate of \citet{DKV_16}.
We emphasize the difference between the explosion and shock breakout moments.
Henceforth the SN age is counted from the adopted explosion moment.
A low luminosity of the pre-SN model before the shock breakout at 2.42\,days
   naturally accounts for the absence of a detectable emission on July 23.54
   \citep{SKS_13}.

\begin{figure}
   \includegraphics[width=\columnwidth, clip, trim=9 148 49 132]{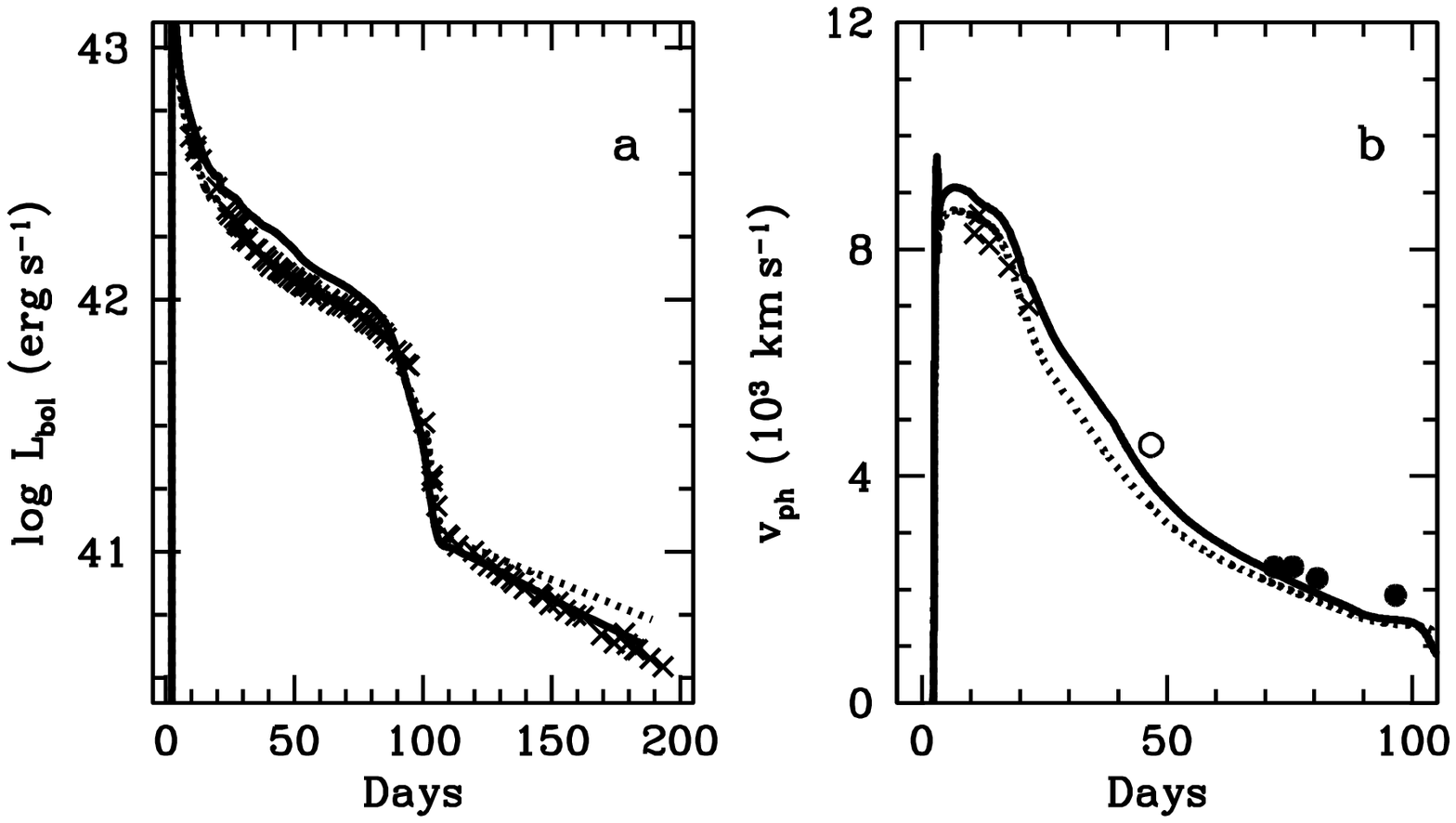}
   \caption{%
   Hydrodynamic models m23e1.2 (\emph{dotted line\/}) and m26e1.4
      (\emph{solid line\/}) for SN~2013ej.
   Panel \textbf{a}: the calculated bolometric light curves are overplotted
      on the bolometric data (\emph{crosses\/}) reported by \citet{DKV_16}.
   Panel \textbf{b}: the calculated photospheric velocity is compared with
      photospheric velocities estimated from the absorption minimum of
      the \ion{Si}{ii}~6355\,\AA\ line (\emph{crosses\/}) by \citet{DKV_16}
      and recovered from the \ion{Na}{i} doublet (\emph{open circle\/}) and
      \ion{Fe}{ii}~6148\,\AA\ (\emph{filled circles\/}) profiles.
   }
   \label{fig:blcvph}
\end{figure}

Model m23e1.2 with the low amount (0.02\,$M_{\sun}$) and moderate mixing
   of radioactive $^{56}$Ni fits the bolometric light curve at the
   plateau stage (Fig.~\ref{fig:blcvph}a), but fails at the radioactive
   tail where the calculated light curve declines slower compared to the
   observations.
The latter is directly related to the deep location of $^{56}$Ni.
The problem is solved in model m26e1.4 with the larger mass and more
   extended distribution $^{56}$Ni (Table~\ref{tab:hydmod}).
The radial distribution of $^{56}$Ni in this case (Fig.~\ref{fig:denni})
   is a spherical representation of the bipolar $^{56}$Ni jets, which are
   recovered below from the H$\alpha$ line and the radioactive tail.
Solving the problem of the radioactive tail, model m26e1.4, however, shows 
   the 20\% larger luminosity at the photospheric stage.
The excess stems from the spherical distribution of $^{56}$Ni in the model,
   which ignores the fact that the bulk of $^{56}$Ni resides in the rear
   hemisphere (see Section~\ref{sec:disc}).
With these reservations model m26e1.4 is more preferred compared to m23e1.2.

Along with the bolometric light curve, the velocity at the photosphere level
   (photospheric velocity, for short) is another observable that should be met
   to constrain the hydrodynamic model.
The photospheric velocity can be recovered from either absorption minima of
   weak metal lines or the profile modelling of stronger isolated lines.
At the early phase we use the photospheric velocities obtained from
   \ion{Si}{ii}~6355\,\AA\ line \citep{DKV_16}. 
A photospheric velocity of 4550\kms\ on day 44%
\footnote{Hereafter the epochs of spectral observations are taken according to
   \citet{DKV_16}}
   is found using the Monte Carlo simulation of \ion{Na}{i} 5890, 5896\,\AA\
   doublet. 
Photospheric velocities of 2400, 2400, 2200, and 1900\kms\ on days 69, 73, 78,
   and 94, respectively, were inferred from the \ion{Fe}{ii}~6148\,\AA\ line
   profile.
The uncertainty of these velocity values does not exceed $\pm100$\kms.
Both hydrodynamic models reproduce the evolution of photospheric velocity
   (Fig.~\ref{fig:blcvph}b) with a marginally better fit in the case of
   model m26e1.4.

\section{Nebular H$\alpha$ line and $^{56}$Ni jets}
\label{sec:hajets}
%
\begin{figure}
   \includegraphics[width=\columnwidth, clip, trim=15 18 44 212]{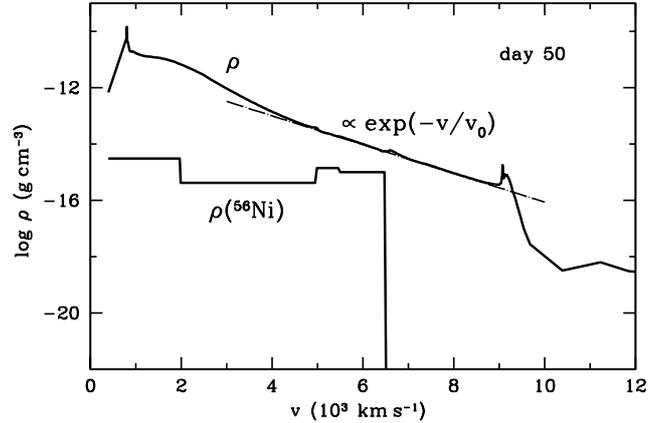}
   \caption{%
   The gas and $^{56}$Ni density as a function of velocity for the fiducial
      model m26e1.4 at $t=50$ days (\emph{solid lines\/}).
   \emph{Dash-dotted line\/} is the exponential fit $\exp(-v/v_0)$.
   }
   \label{fig:denni}
\end{figure}
The broad appearance and pronounced asymmetry of the nebular H$\alpha$ emission
   suggests that the ejecta likely harbors bipolar $^{56}$Ni ejecta
   \citep{BSK_15} likewise in the type IIP SN~2004dj \citep{CFS_05}.
Here we explore this conjecture for SN~2013ej with the intention to infer
   parameters of the $^{56}$Ni distribution from the H$\alpha$ profile and
   the radioactive tail of the bolometric light curve.

The exponential density distribution $\rho \propto \exp(-v/v_0)$ is assumed
   for the ejecta mass of 26\,$M_{\sun}$ and the kinetic energy of
   $1.4\times10^{51}$\,erg, in line with the fiducial model m26e1.4. 
Note that the exponential density distribution fits the results of 
   hydrodynamic modelling (Fig.~\ref{fig:denni}).
The hydrogen abundance $X = 0.7$ is assumed to be uniform throughout the ejecta.
The $^{56}$Ni distribution is set by a central spherical component with the
   mass $M_s$, the outer velocity $v_s$, and the bipolar conical ejecta (jets)
   with the half-opening angle $\psi$ and the inclination angle $\theta$;
   in other respect the jets may be different.
In the near hemisphere the jet (dubbed ``blue jet'') contains $M_b$ of
   $^{56}$Ni and lies in the velocity range $v_{b1} < v < v_{b2}$; 
   the ``red jet'' in the rear hemisphere contains $M_r$ of $^{56}$Ni and
   lies in the velocity range $v_{r1} < v < v_{r2}$.
The $^{56}$Ni density of each component is assumed to be uniform to minimize 
   a number of free parameters. 
The $^{56}$Ni presumably does not affect the background ejecta density.

Parameters of the $^{56}$Ni distribution are constrained by the two major
   observables: the nebular H$\alpha$ profile and the luminosity evolution 
   of the radioactive tail; both sets of data are taken from \citet{DKV_16}. 
The distribution of the energy deposition throughout the ejecta is calculated
   in a single flight approximation for gamma-quanta with the absorption
   coefficient $k_{\gamma} = 0.06 Y_e$, where $Y_{e}$ is a number of electrons
   per nucleon \citep{KF_92}.
Positrons from the $^{56}$Co decay presumably deposit their energy on-the-spot
   with the annihilation quanta taken into account in the gamma-ray emission
   of $^{56}$Co decay.
The deposited energy is assumed to transform into radiation instantly.
This approximation is valid until $\sim$900\,days when a freeze out effect
   gets noticeable \citep{FK_93}.
The H$\alpha$ emissivity is assumed to be proportional to the local deposition
   rate $\epsilon$.
This assumption is sufficient to obtain the H$\alpha$ profile in relative
   fluxes.
However, to take into account Thomson scattering effects in the emission
   profiles the distribution of electron concentration is also needed.

The local electron number density $n_e$ can be determined from the ionization
   balance $\epsilon/w = \alpha n_e^2$, where $w$ is the work per one hydrogen
   ionization, and $\alpha$ is the recombination coefficient for upper levels
   $n>2$, i.e., for the recombination case C, appropriate at the early nebular
   stage of SNe~IIP; the electron temperature of 5000\,K is assumed.
The standard value $w = 36$\,eV was obtained for the collisional ionization
   losses of fast electrons in the hydrogen \citep{DYL_99}.
In SNe~IIP a significant amount of deposited energy goes into the ultraviolet
   radiation that is able reionize hydrogen from the second level thus reducing
   the $w$ value.
One can use observational data of SN~1987A at the nebular epoch to infer
   a modified value of $w$.
At the nebular stage, each hydrogen recombination creates one H$\alpha$
   photon, so if the H$\alpha$ quanta escape, then the H$\alpha$ luminosity
   is related to the bolometric luminosity as 
   $L(\mathrm{H}\alpha) = (h\nu/w) L_{bol}$ (here $h\nu$ is the energy of the
   H$\alpha$ photon).
This ratio is violated at the early nebular stage, when the H$\alpha$ radiation
   is affected by an absorption.
In SN~1987A the H$\alpha$ is saturated at $t < 200$\,days which is indicated
   by the plateau in the H$\alpha$ light curve during $150-200$\,days
   \citep{Han_91}.
On day 200, when the saturation effect becomes negligible, the SN~1987A 
   H$\alpha$ and bolometric luminosities suggest
   $L(\mathrm{H}\alpha) \approx 0.1 L_{bol}$ \citep{Han_91}, which in turn
   implies $w \approx 20$\,eV; this value is used below.

\begin{table}
\centering
\caption{Parameters of $^{56}$Ni jets.}
\label{tab:nijets}
\begin{tabular}{lcc}
\hline
\noalign{\smallskip}
 Parameter &  Symbol & Value \\
\noalign{\smallskip}
\hline 
\noalign{\smallskip}
Jets inclination          &      $\theta$  &  56$^{\circ}$ \\
Half-opening angle         &     $\psi$    &  30$^{\circ}$ \\
Mass of spherical component &    $M_s$     &  0.004\,$M_{\sun}$ \\
Mass of blue jet             &   $M_b$     &  0.011\,$M_{\sun}$ \\ 
Mass of red jet              &   $M_r$     &  0.024\,$M_{\sun}$ \\ 
Velocity of spherical component & $v_s$    &  2000\kms \\
Inner boundary of blue jet  &    $v_{b1}$  &  2000\kms \\
Outer boundary of blue jet  &    $v_{b2}$  &  5500\kms \\
Inner boundary of red jet   &    $v_{r1}$  &  5000\kms \\
Outer boundary of red jet   &    $v_{r2}$  &  6500\kms \\
\noalign{\smallskip}
\hline
\end{tabular}
\end{table}
\begin{figure}
   \includegraphics[width=\columnwidth, clip, trim=8 18 26 82]{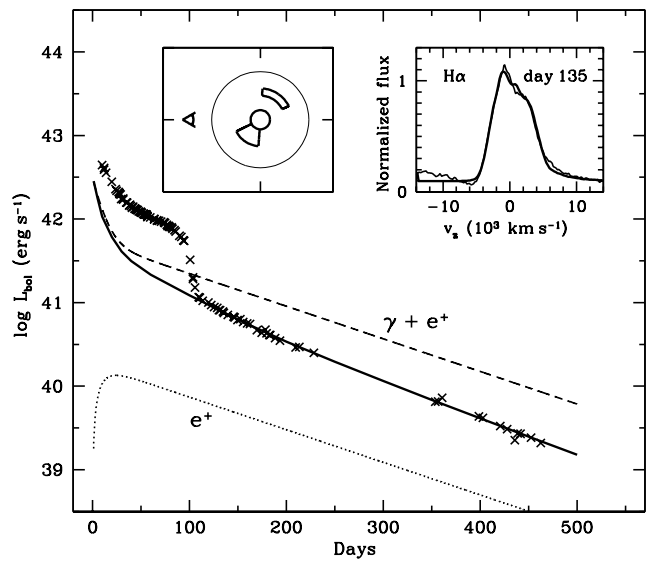}
   \caption{%
   Bipolar $^{56}$Ni jet model (see Table~\ref{tab:nijets}) for the radioactive
      tail and the nebular H$\alpha$ profile.
   \emph{Solid line\/} is the model luminosity deposited by radioactive decay, 
      \emph{dashed line\/} is the total luminosity released by radioactive
      decay, \emph{dotted line\/} is the power deposited by positrons, and
      crosses are the bolometric data reported by \citet{DKV_16}.
   Inset in the left upper corner shows the model configuration of $^{56}$Ni
      components with respect to observer (on the left); the circle
      depicts the level in the ejecta at the velocity of 10\,000\kms.
   Right inset shows the calculated H$\alpha$ (\emph{thick solid line\/})
      compared to that observed by \citeauthor{DKV_16} on day 135
      (\emph{thin solid line\/}).
   }
   \label{fig:jetsmod}
\end{figure}
The optimal set of parameters of the $^{56}$Ni ejecta is given in
   Table~\ref{tab:nijets}.
The H$\alpha$ profile and the radioactive tail of the bolometric light curve
   are described fairly well with the found model of the bipolar $^{56}$Ni
   ejecta (Fig.~\ref{fig:jetsmod}).
Remarkably, the model accounts for both the fast decline after day 110 and
   the slowdown of decline after day 200, the fact emphasized by
   \citet{DKV_16}.
In our search of the optimal parameters we bound ourselves with the requirement
   of the minimal $^{56}$Ni mass consistent with the H$\alpha$ profile and
   the radioactive tail.
The upper limit of the $^{56}$Ni mass cannot be reliably inferred by this kind
   of modelling because a high-velocity $^{56}$Ni ($v>7000$\kms) does not
   affect significantly the H$\alpha$ profile and the radioactive tail at
   the nebular stage.

\section{$^{56}$Ni jets and polarization}
\label{sec:polar}
%
\begin{figure}
   \includegraphics[width=\columnwidth, clip, trim=22 21 52 208]{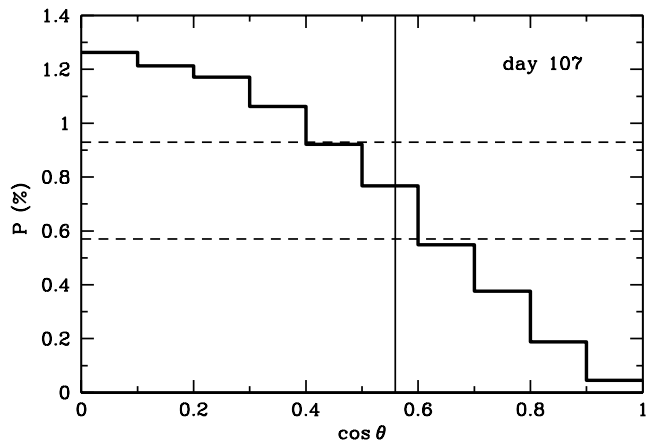}
   \caption{%
   Linear polarization on day 107 as function of cosine of the jet inclination
      angle for the adopted model (\emph{thick solid line\/})
      (see Table~\ref{tab:nijets}). 
   \emph{Vertical line\/} marks the optimal inclination for the jet model,
      \emph{dashed lines\/} show the observational range from two choices
      of interstellar polarization \citep{MDJ_17}. 
   }
   \label{fig:linpol}
\end{figure}
The aspherical ionization pattern produced by the $^{56}$Ni jets should
   result in the linear polarization due to the Thomson scattering likewise
   in SN~2004dj \citep{Chu_06}.
The question arises, whether the polarization in our model is able to account
   for the observed polarization in SN~2013ej \citep{MDJ_17,KPEK_16}.
We consider only the nebular phase to minimize effects of the radiation
   transfer in the opaque ejecta.
The polarization is calculated using the Monte Carlo technique \citep{Chu_06}.
The model suggests that a quasi-continuum optical radiation is emitted with
   the rate proportional to the local deposition rate, the approximation 
   valid at the nebular phase. 

The computed polarization for the adopted ejecta and the $^{56}$Ni jets model
   on day 107 is plotted in Fig.~\ref{fig:linpol} as a function of the cosine
   of the inclination angle. 
For the optimal inclination angle $\theta = 56^{\circ}$ the model polarization
   lies in the range between the observed polarization values obtained 
   with alternative assumptions about the interstellar polarization
   \citep{MDJ_17}.
The plot demonstrates that our model predicts the linear polarization 
   consistent with the observational data. 
We conclude therefore that most (if not all) of the observed polarization
   seems to be related to the bipolar $^{56}$Ni ejecta.

A doubt may arise concerning the assumed collinearity of $^{56}$Ni jets.
It may well be that the $^{56}$Ni ejecta could form as a result of the
   RT instability caused by the shock propagation \citep[e.g.][]{WMJ_15}.
In this case the opposite $^{56}$Ni ejecta generally are not collinear.
Although our model of collinear jets agrees with the H$\alpha$ and polarization
  data, we admit that this does not rule out the jets non-collinearity.

\section{Discussion and Conclusions}
\label{sec:disc}
Our aim has been to recover parameters of the hydrodynamic model for the
   unusual type IIP SN~2013ej and to explore the effects of a possible
   asymmetry of the $^{56}$Ni ejecta.
We failed to find a unique spherical hydrodynamic model that would be able
   to fit well the light curve at both stages, the plateau and the radioactive
   tail.
The model with $0.02\,M_{\sun}$ of the center-concentrated $^{56}$Ni describes
   the photospheric stage fairly well, but fails at the radioactive tail.
The model with the twice as large amount of $^{56}$Ni residing at high
   velocities excellently fits the radioactive tail, but overproduces the
   luminosity at the plateau by a factor of $\approx$1.2.
Yet this model is preferred because the latter mismatch stems from the
   spherical representation of the bipolar $^{56}$Ni jets.
In the real 3D picture with the bulk of radioactive $^{56}$Ni residing in
   the far hemisphere, the flux from the near hemisphere should be lower
   than in the spherical model due to the occultation effect. 
A correct description of the light curve of SN~2013ej at the photospheric
   stage therefore requires a multi-dimensional radiation hydrodynamics.

\begin{table}
\centering
\caption[]{Hydrodynamic models of type IIP supernovae.}
\label{tab:sumtab}
\begin{tabular}{@{ } l  c  c @{ } c @{ } c @{ } c  c @{ }}
\hline
\noalign{\smallskip}
 SN & $R_0$ & $M_{ej}$ & $E$ & $M_{\mathrm{Ni}}$ 
       & $v_{\mathrm{Ni}}^{max}$ & $v_{\mathrm{H}}^{min}$ \\
       & $(R_{\sun})$ & $(M_{\sun})$ & ($10^{51}$ erg) & $(10^{-2} M_{\sun})$
       & \multicolumn{2}{c}{(km\,s$^{-1}$)}\\
\noalign{\smallskip}
\hline
\noalign{\smallskip}
 1987A &  35  & 18   & 1.5    & 7.65 &  3000 & 600 \\
1999em & 500  & 19   & 1.3    & 3.6  &  660  & 700 \\
2000cb &  35  & 22.3 & 4.4    & 8.3  &  8400 & 440 \\
 2003Z & 230  & 14   & 0.245  & 0.63 &  535  & 360 \\
2004et & 1500 & 22.9 & 2.3    & 6.8  &  1000 & 300 \\
2005cs & 600  & 15.9 & 0.41   & 0.82 &  610  & 300 \\
2008in & 570  & 13.6 & 0.505  & 1.5  &  770  & 490 \\
2009kf & 2000 & 28.1 & 21.5   & 40.0 &  7700 & 410 \\
2012A  &  715 & 13.1 & 0.525  & 1.16 &  710  & 400 \\
2013ej & 1500 & 26.1 & 1.4    & 3.9  &  6500 & 800 \\
\noalign{\smallskip}
\hline
\end{tabular}
\end{table}
The hydrodynamic modelling of SN~2013ej lead us to conclude that the ejecta
   mass is $23.1-26.1\,M_{\sun}$, the explosion energy is
   $(1.18-1.40)\times10^{51}$\,erg, the $^{56}$Ni mass is
   $0.020-0.039\,M_{\sun}$, with the upper limits being preferred. 
The pre-SN should be a very extended RSG star with the radius of
   1500\,$R_{\sun}$.
The large pre-SN radius in our model is responsible for the luminous broad
   initial peak of the light curve.
It is notable that the circumstellar (CS) shell with the radius of
   $1300-1500\,R_{\sun}$ and mass of $\sim$1\,$M_{\sun}$ invoked by
   \citet{MPV_17} can be considered as a proxy for our extended RSG model.
Other SN parameters of both models cannot be meaningfully compared, because
   the information on the photospheric velocities in the model of
   \citet{MPV_17} is lacking.  
Yet we note that our ejecta mass is twice as large.

It is instructive to confront the parameters of SN~2013ej with those of
   another luminous type IIP SN~2004et (Table~\ref{tab:sumtab}).
Both have similar ejecta masses and pre-SN radii.
However, the explosion energy and the $^{56}$Ni mass of SN~2013ej are by
   a factor of $\sim$1.6 lower than those of SN~2004et. 
Moreover, the maximal velocities of $^{56}$Ni are dramatically different: 
   1000\kms\ in SN~2004et vs. 6500\kms\ in SN~2013ej.
This disparity indicates that the explosion outcome in SNe~IIP can be
   significantly different even for comparable pre-SN masses.

An intriguing point is the physics behind the steep decline of the light curve
   of SN~2013ej at the photospheric stage.
While the initial broad luminosity peak postpones the cooling and recombination
   wave regime, this does not explain the lack of the flat plateau at the later
   photospheric stage.
In contrast, SN~2004et after about day 35 shows a {\em flat} plateau 
   \citep{SASM_06}.  
The reason for the different behavior of SN~2013ej at this stage is hidden in
   the density distribution of a pre-SN model.
Specifically, at the transition between the helium core and the hydrogen 
   envelope the density gradient in the pre-SN model of SN~2013ej varies with
   the radius more smoothly compared to that of SN~2004et. 
We attribute this distinction to the different outcome of the RT mixing
   during the explosion, viz., in SN~2013ej mixing was more vigorous than
   in SN~2004et.
This conjecture seems to be supported by the low velocities of the ${56}$Ni
   matter in SN~2004et ($v_{\mathrm{Ni}}^{max} = 1000$\kms) and the high
   ${56}$Ni velocities in SN~2013ej ($v_{\mathrm{Ni}}^{max} = 6500$\kms).
Whether the proposed explanation of SN~IIP/L phenomena is universal remains
   to be verified.

\begin{figure}
   \includegraphics[width=\columnwidth, clip, trim=0 23 27 27]{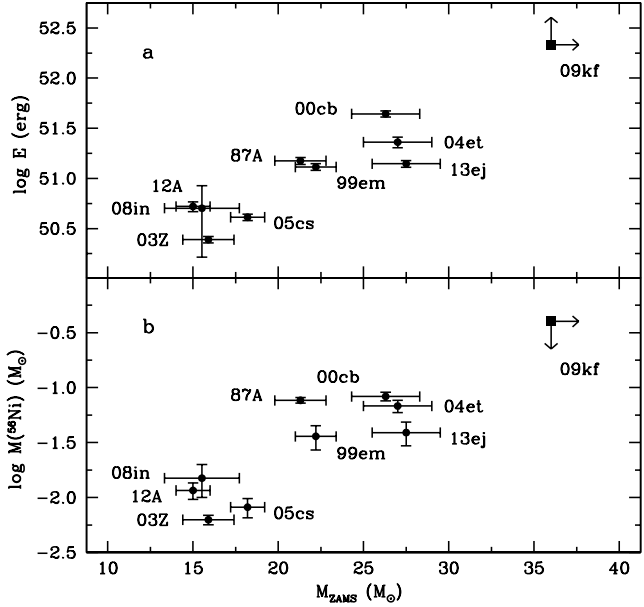}
   \caption{%
   Explosion energy (Panel \textbf{a}) and $^{56}$Ni mass (Panel \textbf{b})
      vs. hydrodynamic progenitor mass for SN~2013ej and nine other
      core-collapse SNe \citep{UC_15}.
   }
   \label{fig:ennims}
\end{figure}

Combining the SN~2013ej ejecta mass with a neutron star of 1.4\,$M_{\sun}$
   and a moderate mass of $1-2\,M_{\sun}$ lost by the stellar wind
   \citep[cf.][]{UC_09}, one comes to the progenitor mass of
   $25.5-29.5\,M_{\sun}$.
The parameters of our fiducial model m26e1.4 on the diagrams of the explosion
   energy vs. the progenitor mass and the $^{56}$Ni mass vs. the progenitor
   mass (Fig.~\ref{fig:ennims}) place SN~2013ej into the high mass region
   at the lower boundary of both scatter plots.  
Now the sample of well-observed SNe~IIP studied hydrodynamically in a uniform
   way \citep{UC_15} mounts up to ten and this reinforces our former impression
   that the SN~IIP mass distribution is skewed towards high masses with an
   apparent deficit of SNe~IIP in the range of $9-15\,M_{\sun}$.
This in turn brings about a tension with general wisdom that stars with the
   main-sequence masses of $9-25\,M_{\sun}$ produce SNe~II.
The tension is alleviated, if one admits that that SNe from the mass range
   of $9-15\,M_{\sun}$ are very faint, so they escape detection.
This serious issue requires a further thorough study.

The model of extended asymmetric $^{56}$Ni ejecta of SN~2013ej is found to 
   be compatible with the polarization data at the early nebular epoch.
\citet{MDJ_17} suggest that the Thomson scattering in an oblate ellipsoidal
   envelope and the dust scattering in the CS envelope could be responsible
   for the polarization in SN~2013ej.
In this regard we notice that the mechanism of the asymmetric ionization
   related to the CS interaction invoked by \citet{MDJ_17}, cannot noticeably
   modify the asymmetry produced by the bipolar $^{56}$Ni ejecta.
Indeed, the X-ray luminosity associated with the CS interaction on day 114 is
   three orders lower than the bolometric luminosity at the same epoch
   \citep[cf.][]{CRS_16}.

The bipolar structure of $^{56}$Ni in SN~2013ej is reminiscent of
   SN~2004dj \citep{CFS_05} with an exception that the $^{56}$Ni amount in
   SN~2013ej is twice as large and the maximal velocity of $^{56}$Ni is
   also a factor of two larger.
The high velocity of the $^{56}$Ni ejecta in SN~2013ej is a challenge for 
   the explosion mechanism. 
In the neutrino-driven explosion mechanism, high velocities of the
   $^{56}$Ni-rich matter in SNe~IIP are explained as an outcome of the RT
   instability that accompanies the shock wave propagation in the inner 
   parts of the exploding star.
Recent 3D simulations of an explosion of a 15\,$M_{\sun}$ RSG model demonstrate
   that the RT plumes of $^{56}$Ni are able to protrude up to $4000-5000$\kms\
   \citep{WMJ_15}.
However, the simulations show also that in a more massive star of
   20\,$M_{\sun}$ the $^{56}$Ni velocities are a factor of 1.5 lower.
Thus, at the moment it is not clear, whether the RT mechanism is able to
   account for the $^{56}$Ni velocities of $\sim$6000\kms\ in SNe with
   the mass of $\sim$25\,$M_{\sun}$.
It is noteworthy that the velocity extent of $^{56}$Ni in SN~2013ej is not
   extreme among SNe~IIP: in SN~2000cb the $^{56}$Ni matter seems to be
   mixed up to 8400\kms\ \citep{UC_11}.
In this regard it may well be that apart from the RT mixing some additional
   asymmetry of the explosion could be involved in the buildup of high
   velocities of $^{56}$Ni.
One way or another, the asymmetry and high velocities of the $^{56}$Ni ejecta
   in SNe~IIP should be considered as crucial observational constraints for
   the explosion mechanism.
Finally, note that the bipolar high-velocity $^{56}$Ni ejecta in SN~2013ej
   is not unique phenomenon: the well-known high-velocity bipolar jets in
   Cas A (SN~IIb) are the firmly established feature that apparently has
   an explosion origin \citep{FM_16}.

For SN~2013ej our estimate of the progenitor mass (dubbed ``hydrodynamic
   mass'') is significantly larger than $8-15.5$\,$M_{\sun}$ (dubbed
   ``pre-SN luminosity mass'') inferred from pre-explosion {\em HST}
   $F814W$-filter image \citep{FMS_14}.
The SN~IIP progenitor mass problem was recovered and discussed first
   in the context of SN~2005cs \citep{UC_08}, in which case
   hydrodynamic mass turned out to be significantly larger than the ZAMS
   mass from archival image \citep{MSD_05}.
Since then the mass problem is reproduced {\em every time} when both mass
   estimates are available.
The exception is SN~1987A, for which both methods provide similar mass
   estimates of $\sim$20\,$M_{\sun}$.
This in turn indicates that the mass problem arises possibly only in cases
   when the SN~IIP event is related to the RSG explosion. 
 
The conflict between hydrodynamic and pre-SN luminosity masses suggests that
   either the former or latter method is in error. 
Our hydrodynamic code in case of SN~1999em --- classic SN~IIP --- produces 
   similar SN parameter values \citep{Utr_07} to those obtained by independent
   codes of \citet{BBP_05} and \citet{BBH_11}.
It should be emphasized that the codes use different treatment of radiation
   transport: one-group radiation transfer \citep{Utr_07}, multi-group
   radiation transfer \citep{BBP_05}, and flux-limited equilibrium
   diffusion approximation \citep{BBH_11}.
This indicates the robustness of the recovered parameter values provided that
   the observables (bolometric light curve and photospheric velocities) are
   adequately modeled.
It should be highlighted that in some cases hydrodynamic studies of SNe~IIP
   are limited to the light curve modelling, ignoring photospheric velocities.
In this approach the ejecta mass and the explosion energy cannot be reliably
   determined.
It is equally dangerous to use approximation formulae for the parameter
   estimates, since such an approach ignores the early behavior of the
   luminosity and expansion velocity, which contain an additional information
   on the ejecta mass, explosion energy, and pre-SN radius.

Until recently a sensitive point of the hydrodynamic approach was a ``manual''
   composition mixing and density smoothing between the metal core, the helium
   core, and the hydrogen envelope of a pre-SN star, although that was well
   constrained by the light curve at the late plateau.
The justification for the manual mixing is the inability of 1D hydrodynamics
   to treat the Rayleigh-Taylor instability and mixing initiated by the shock
   propagation in the pre-SN.
The recent study dedicated to mixing issue on the basis of the 3D RSG explosion
   \citep{UWJM_17} shows that the manual mixing applied earlier in
   1D model of SN~1999em \citep{Utr_07} is fully consistent with the mixing
   produced by the 3D hydrodynamics.  
At the moment we do not see another way to seriously improve our hydrodynamic
   model.

Alternatively, the pre-SN luminosity mass could be underestimated.
The most apparent reason is that the pre-SN light might be obscured by a dusty
   CS envelope \citep{UC_08, SECM_09}.
This possibility finds some support in the growing number of SNe~IIP with
   signatures of a confined dense CS envelope in early spectra, e.g., SN~2006bp
   \citep{QWH_07} and SN~2013fs \citep{YPGY_17}.
These data indicate a dense CS shell at the radii $\la$10$^{15}$\,cm.
To illustrate the point, let us estimate possible optical depth of the CS shell
   of the mass $M_s$ at the radius $R_s = 10^{15}$\,cm.
The equilibrium temperature of the shell for the stellar luminosity of
   $3 \times 10^5~L_{\sun}$, characteristic of a 25\,$M_{\sun}$ RSG star, is
   then $\sim$900\,K, i.e., below the silicate dust condensation temperature
   ($1000-1200$\,K) for the relevant gas pressure.
We adopt the standard dust-to-gas ratio of $10^{-2}$ and the optical properties
   of a silicate dust \citep{DL_84}, which suggest the absorption efficiency
   $Q_a = 0.4(a/\lambda)$ for the grain radius $a$ of about $10^{-5}$\,cm.
The optical depth of the CS shell is then
  $\tau = 2.2(M_s/10^{-3}M_{\sun})(R_s/10^{15}\mbox{cm})^{-2}(0.8\mu\mbox{m}/\lambda)$. 
The estimate shows that the shell with the mass $M_s \sim 10^{-3}M_{\sun}$, 
   produced by a sensible RSG mass-loss rate, could maintain significant
   absorption of the pre-SN light. 

Another independent progenitor mass estimate for SN~2013ej ($12-15$\,$M_{\sun}$)
   is obtained from the nebular oxygen doublet [O\,I] 6300, 6364\,\AA\
   \citep{YJV_16}.
The accuracy of this method depends on the uncertainty of the adopted density,
   $^{56}$Ni distribution, and molecules (i.e., CO and SiO) abundance of
   the O-rich matter.
Note that the cooling by molecules is ignored in the current model.
Furthermore, the adopted $^{56}$Ni distribution set by centrally concentrated
   sphere is an oversimplification in the case of SN~2013ej.
Therefore, although this analysis of nebular spectra is an interesting
   alternative approach, there remain uncertainties that cast shadow on
   the reliability of the progenitor mass estimate.

\section*{Acknowledgements}
We thank Govinda Dhungana for kindly sending us the spectra of SN~2013ej. 
The study by N.C. of $^{56}$Ni asymmetry is partially supported by Russian
   Scientific Foundation grant 16-12-10519.


\bsp	
\label{lastpage}

\begin{thebibliography}{}
\makeatletter
\relax
\def\mn@urlcharsother{\let\do\@makeother \do\$\do\&\do\#\do\^\do\_\do\%\do\~}
\def\mn@doi{\begingroup\mn@urlcharsother \@ifnextchar [ {\mn@doi@}
  {\mn@doi@[]}}
\def\mn@doi@[#1]#2{\def\@tempa{#1}\ifx\@tempa\@empty \href
  {http://dx.doi.org/#2} {doi:#2}\else \href {http://dx.doi.org/#2} {#1}\fi
  \endgroup}
\def\mn@eprint#1#2{\mn@eprint@#1:#2::\@nil}
\def\mn@eprint@arXiv#1{\href {http://arxiv.org/abs/#1} {{\tt arXiv:#1}}}
\def\mn@eprint@dblp#1{\href {http://dblp.uni-trier.de/rec/bibtex/#1.xml}
  {dblp:#1}}
\def\mn@eprint@#1:#2:#3:#4\@nil{\def\@tempa {#1}\def\@tempb {#2}\def\@tempc
  {#3}\ifx \@tempc \@empty \let \@tempc \@tempb \let \@tempb \@tempa \fi \ifx
  \@tempb \@empty \def\@tempb {arXiv}\fi \@ifundefined
  {mn@eprint@\@tempb}{\@tempb:\@tempc}{\expandafter \expandafter \csname
  mn@eprint@\@tempb\endcsname \expandafter{\@tempc}}}

\bibitem[\protect\citeauthoryear{{Baklanov}, {Blinnikov}  \&
  {Pavlyuk}}{{Baklanov} et~al.}{2005}]{BBP_05}
{Baklanov} P.~V.,  {Blinnikov} S.~I.,   {Pavlyuk} N.~N.,  2005, \mn@doi
  [Astronomy Letters] {10.1134/1.1958107}, \href
  {http://adsabs.harvard.edu/abs/2005AstL...31..429B} {31, 429}

\bibitem[\protect\citeauthoryear{{Bersten}, {Benvenuto}  \& {Hamuy}}{{Bersten}
  et~al.}{2011}]{BBH_11}
{Bersten} M.~C.,  {Benvenuto} O.,   {Hamuy} M.,  2011, \mn@doi [\apj]
  {10.1088/0004-637X/729/1/61}, \href
  {http://adsabs.harvard.edu/abs/2011ApJ...729...61B} {729, 61}

\bibitem[\protect\citeauthoryear{{Bose} et~al.,}{{Bose} et~al.}{2015}]{BSK_15}
{Bose} S.,  et~al., 2015, \mn@doi [\apj] {10.1088/0004-637X/806/2/160}, \href
  {http://adsabs.harvard.edu/abs/2015ApJ...806..160B} {806, 160}

\bibitem[\protect\citeauthoryear{{Chakraborti} et~al.,}{{Chakraborti}
  et~al.}{2016}]{CRS_16}
{Chakraborti} S.,  et~al., 2016, \mn@doi [\apj] {10.3847/0004-637X/817/1/22},
  \href {http://adsabs.harvard.edu/abs/2016ApJ...817...22C} {817, 22}

\bibitem[\protect\citeauthoryear{{Chugai}}{{Chugai}}{2006}]{Chu_06}
{Chugai} N.~N.,  2006, \mn@doi [Astronomy Letters] {10.1134/S1063773706110041},
  \href {http://adsabs.harvard.edu/abs/2006AstL...32..739C} {32, 739}

\bibitem[\protect\citeauthoryear{{Chugai}, {Fabrika}, {Sholukhova},
  {Goranskij}, {Abolmasov}  \& {Vlasyuk}}{{Chugai} et~al.}{2005}]{CFS_05}
{Chugai} N.~N.,  {Fabrika} S.~N.,  {Sholukhova} O.~N.,  {Goranskij} V.~P.,
  {Abolmasov} P.~K.,   {Vlasyuk} V.~V.,  2005, \mn@doi [Astronomy Letters]
  {10.1134/1.2138766}, \href
  {http://adsabs.harvard.edu/abs/2005AstL...31..792C} {31, 792}

\bibitem[\protect\citeauthoryear{{Dalgarno}, {Yan}  \& {Liu}}{{Dalgarno}
  et~al.}{1999}]{DYL_99}
{Dalgarno} A.,  {Yan} M.,   {Liu} W.,  1999, \mn@doi [\apjs] {10.1086/313267},
  \href {http://adsabs.harvard.edu/abs/1999ApJS..125..237D} {125, 237}

\bibitem[\protect\citeauthoryear{{Dhungana} et~al.,}{{Dhungana}
  et~al.}{2016}]{DKV_16}
{Dhungana} G.,  et~al., 2016, \mn@doi [\apj] {10.3847/0004-637X/822/1/6}, \href
  {http://adsabs.harvard.edu/abs/2016ApJ...822....6D} {822, 6}

\bibitem[\protect\citeauthoryear{{Draine} \& {Lee}}{{Draine} \&
  {Lee}}{1984}]{DL_84}
{Draine} B.~T.,  {Lee} H.~M.,  1984, \mn@doi [\apj] {10.1086/162480}, \href
  {http://adsabs.harvard.edu/abs/1984ApJ...285...89D} {285, 89}

\bibitem[\protect\citeauthoryear{{Fesen} \& {Milisavljevic}}{{Fesen} \&
  {Milisavljevic}}{2016}]{FM_16}
{Fesen} R.~A.,  {Milisavljevic} D.,  2016, \mn@doi [\apj]
  {10.3847/0004-637X/818/1/17}, \href
  {http://adsabs.harvard.edu/abs/2016ApJ...818...17F} {818, 17}

\bibitem[\protect\citeauthoryear{{Fransson} \& {Kozma}}{{Fransson} \&
  {Kozma}}{1993}]{FK_93}
{Fransson} C.,  {Kozma} C.,  1993, \mn@doi [\apjl] {10.1086/186822}, \href
  {http://adsabs.harvard.edu/abs/1993ApJ...408L..25F} {408, L25}

\bibitem[\protect\citeauthoryear{{Fraser} et~al.,}{{Fraser}
  et~al.}{2014}]{FMS_14}
{Fraser} M.,  et~al., 2014, \mn@doi [\mnras] {10.1093/mnrasl/slt179}, \href
  {http://adsabs.harvard.edu/abs/2014MNRAS.439L..56F} {439, L56}

\bibitem[\protect\citeauthoryear{{Hanuschik}}{{Hanuschik}}{1991}]{Han_91}
{Hanuschik} R.~W.,  1991, in {Danziger} I.~J.,  {Kjaer} K.,  eds,  European
  Southern Observatory Conference and Workshop Proceedings Vol. 37, European
  Southern Observatory Conference and Workshop Proceedings. p.~237

\bibitem[\protect\citeauthoryear{{Huang} et~al.,}{{Huang}
  et~al.}{2015}]{HWZ_15}
{Huang} F.,  et~al., 2015, \mn@doi [\apj] {10.1088/0004-637X/807/1/59}, \href
  {http://adsabs.harvard.edu/abs/2015ApJ...807...59H} {807, 59}

\bibitem[\protect\citeauthoryear{{Kim} et~al.,}{{Kim} et~al.}{2013}]{KZL_13}
{Kim} M.,  et~al., 2013, Central Bureau Electronic Telegrams, \href
  {http://adsabs.harvard.edu/abs/2013CBET.3606....1K} {3606}

\bibitem[\protect\citeauthoryear{{Kozma} \& {Fransson}}{{Kozma} \&
  {Fransson}}{1992}]{KF_92}
{Kozma} C.,  {Fransson} C.,  1992, \mn@doi [\apj] {10.1086/171311}, \href
  {http://adsabs.harvard.edu/abs/1992ApJ...390..602K} {390, 602}

\bibitem[\protect\citeauthoryear{{Kumar}, {Pandey}, {Eswaraiah}  \&
  {Kawabata}}{{Kumar} et~al.}{2016}]{KPEK_16}
{Kumar} B.,  {Pandey} S.~B.,  {Eswaraiah} C.,   {Kawabata} K.~S.,  2016,
  \mn@doi [\mnras] {10.1093/mnras/stv2720}, \href
  {http://adsabs.harvard.edu/abs/2016MNRAS.456.3157K} {456, 3157}

\bibitem[\protect\citeauthoryear{{Lee} et~al.,}{{Lee} et~al.}{2013}]{LLW_13}
{Lee} M.,  et~al., 2013, The Astronomer's Telegram, \href
  {http://adsabs.harvard.edu/abs/2013ATel.5466....1L} {5466}

\bibitem[\protect\citeauthoryear{{Mauerhan} et~al.,}{{Mauerhan}
  et~al.}{2017}]{MDJ_17}
{Mauerhan} J.~C.,  et~al., 2017, \mn@doi [\apj] {10.3847/1538-4357/834/2/118},
  \href {http://adsabs.harvard.edu/abs/2017ApJ...834..118M} {834, 118}

\bibitem[\protect\citeauthoryear{{Maund}, {Smartt}  \& {Danziger}}{{Maund}
  et~al.}{2005}]{MSD_05}
{Maund} J.~R.,  {Smartt} S.~J.,   {Danziger} I.~J.,  2005, \mn@doi [\mnras]
  {10.1111/j.1745-3933.2005.00100.x}, \href
  {http://adsabs.harvard.edu/abs/2005MNRAS.364L..33M} {364, L33}

\bibitem[\protect\citeauthoryear{{Morozova}, {Piro}  \& {Valenti}}{{Morozova}
  et~al.}{2017}]{MPV_17}
{Morozova} V.,  {Piro} A.~L.,   {Valenti} S.,  2017, \mn@doi [\apj]
  {10.3847/1538-4357/aa6251}, \href
  {http://adsabs.harvard.edu/abs/2017ApJ...838...28M} {838, 28}

\bibitem[\protect\citeauthoryear{{Quimby}, {Wheeler}, {H{\"o}flich}, {Akerlof},
  {Brown}  \& {Rykoff}}{{Quimby} et~al.}{2007}]{QWH_07}
{Quimby} R.~M.,  {Wheeler} J.~C.,  {H{\"o}flich} P.,  {Akerlof} C.~W.,  {Brown}
  P.~J.,   {Rykoff} E.~S.,  2007, \mn@doi [\apj] {10.1086/520532}, \href
  {http://adsabs.harvard.edu/abs/2007ApJ...666.1093Q} {666, 1093}

\bibitem[\protect\citeauthoryear{{Sahu}, {Anupama}, {Srividya}  \&
  {Muneer}}{{Sahu} et~al.}{2006}]{SASM_06}
{Sahu} D.~K.,  {Anupama} G.~C.,  {Srividya} S.,   {Muneer} S.,  2006, \mn@doi
  [\mnras] {10.1111/j.1365-2966.2006.10937.x}, \href
  {http://adsabs.harvard.edu/abs/2006MNRAS.372.1315S} {372, 1315}

\bibitem[\protect\citeauthoryear{{Shappee} et~al.,}{{Shappee}
  et~al.}{2013}]{SKS_13}
{Shappee} B.~J.,  et~al., 2013, The Astronomer's Telegram, \href
  {http://adsabs.harvard.edu/abs/2013ATel.5237....1S} {5237}

\bibitem[\protect\citeauthoryear{{Smartt}, {Eldridge}, {Crockett}  \&
  {Maund}}{{Smartt} et~al.}{2009}]{SECM_09}
{Smartt} S.~J.,  {Eldridge} J.~J.,  {Crockett} R.~M.,   {Maund} J.~R.,  2009,
  \mn@doi [\mnras] {10.1111/j.1365-2966.2009.14506.x}, \href
  {http://adsabs.harvard.edu/abs/2009MNRAS.395.1409S} {395, 1409}

\bibitem[\protect\citeauthoryear{{Utrobin}}{{Utrobin}}{2004}]{Utr_04}
{Utrobin} V.~P.,  2004, \mn@doi [Astronomy Letters] {10.1134/1.1738152}, \href
  {http://adsabs.harvard.edu/abs/2004AstL...30..293U} {30, 293}

\bibitem[\protect\citeauthoryear{{Utrobin}}{{Utrobin}}{2007}]{Utr_07}
{Utrobin} V.~P.,  2007, \mn@doi [\aap] {10.1051/0004-6361:20066078}, \href
  {http://adsabs.harvard.edu/abs/2007A%26A...461..233U} {461, 233}

\bibitem[\protect\citeauthoryear{{Utrobin} \& {Chugai}}{{Utrobin} \&
  {Chugai}}{2008}]{UC_08}
{Utrobin} V.~P.,  {Chugai} N.~N.,  2008, \mn@doi [\aap]
  {10.1051/0004-6361:200810272}, \href
  {http://adsabs.harvard.edu/abs/2008A%26A...491..507U} {491, 507}

\bibitem[\protect\citeauthoryear{{Utrobin} \& {Chugai}}{{Utrobin} \&
  {Chugai}}{2009}]{UC_09}
{Utrobin} V.~P.,  {Chugai} N.~N.,  2009, \mn@doi [\aap]
  {10.1051/0004-6361/200912273}, \href
  {http://adsabs.harvard.edu/abs/2009A%26A...506..829U} {506, 829}

\bibitem[\protect\citeauthoryear{{Utrobin} \& {Chugai}}{{Utrobin} \&
  {Chugai}}{2011}]{UC_11}
{Utrobin} V.~P.,  {Chugai} N.~N.,  2011, \mn@doi [\aap]
  {10.1051/0004-6361/201117137}, \href
  {http://adsabs.harvard.edu/abs/2011A%26A...532A.100U} {532, A100}

\bibitem[\protect\citeauthoryear{{Utrobin} \& {Chugai}}{{Utrobin} \&
  {Chugai}}{2015}]{UC_15}
{Utrobin} V.~P.,  {Chugai} N.~N.,  2015, \mn@doi [\aap]
  {10.1051/0004-6361/201424822}, \href
  {http://adsabs.harvard.edu/abs/2015A%26A...575A.100U} {575, A100}

\bibitem[\protect\citeauthoryear{{Utrobin}, {Wongwathanarat}, {Janka}  \&
  {M{\"u}ller}}{{Utrobin} et~al.}{2017}]{UWJM_17}
{Utrobin} V.~P.,  {Wongwathanarat} A.,  {Janka} H.-T.,   {M{\"u}ller} E.,
  2017, \mn@doi [\apj] {10.3847/1538-4357/aa8594}, \href
  {http://adsabs.harvard.edu/abs/2017ApJ...846...37U} {846, 37}

\bibitem[\protect\citeauthoryear{{Wongwathanarat}, {M{\"u}ller}  \&
  {Janka}}{{Wongwathanarat} et~al.}{2015}]{WMJ_15}
{Wongwathanarat} A.,  {M{\"u}ller} E.,   {Janka} H.-T.,  2015, \mn@doi [\aap]
  {10.1051/0004-6361/201425025}, \href
  {http://adsabs.harvard.edu/abs/2015A%26A...577A..48W} {577, A48}

\bibitem[\protect\citeauthoryear{{Woosley}}{{Woosley}}{1988}]{Woo_88}
{Woosley} S.~E.,  1988, \mn@doi [\apj] {10.1086/166468}, \href
  {http://adsabs.harvard.edu/abs/1988ApJ...330..218W} {330, 218}

\bibitem[\protect\citeauthoryear{{Yaron} et~al.,}{{Yaron}
  et~al.}{2017}]{YPGY_17}
{Yaron} O.,  et~al., 2017, \mn@doi [Nature Physics] {10.1038/nphys4025}, \href
  {http://adsabs.harvard.edu/abs/2017NatPh..13..510Y} {13, 510}

\bibitem[\protect\citeauthoryear{{Yuan} et~al.,}{{Yuan} et~al.}{2016}]{YJV_16}
{Yuan} F.,  et~al., 2016, \mn@doi [\mnras] {10.1093/mnras/stw1419}, \href
  {http://adsabs.harvard.edu/abs/2016MNRAS.461.2003Y} {461, 2003}

\makeatother
\end{thebibliography}
\end{document}